\documentclass[
  twocolumn,
  prl,
  amsmath,
  amssymb,
  superscriptaddress,
  floatfix
]{revtex4}

\usepackage{bm}
\usepackage{graphicx}

\begin{document}

\title{Density of states in a two-dimensional chiral metal with vacancies}

\author{P.\ M.\ Ostrovsky}
\affiliation{
 Max Planck Institute for Solid State Research, Heisenbergstr.\ 1, 70569 Stuttgart, Germany
}
\affiliation{
 L.\ D.\ Landau Institute for Theoretical Physics RAS, 119334 Moscow, Russia
}

\author{I.\ V.\ Protopopov}
\affiliation{
 Institut f\"ur Theorie der kondensierten Materie, Karlsruhe Institute of Technology, 76128 Karlsruhe, Germany
}
\affiliation{
 L.\ D.\ Landau Institute for Theoretical Physics RAS, 119334 Moscow, Russia
}

\author{E.\ J.\ K\"onig}
\affiliation{
 Institut f\"ur Theorie der kondensierten Materie, Karlsruhe Institute of Technology, 76128 Karlsruhe, Germany
}

\author{I.\ V.\ Gornyi}
\affiliation{
 Institut f\"ur Nanotechnologie, Karlsruhe Institute of Technology, 76021 Karlsruhe, Germany
}
\affiliation{
 A.\ F.\ Ioffe Physico-Technical Institute, 194021 St.\ Petersburg, Russia.
}

\author{A.\ D.\ Mirlin}
\affiliation{
 Institut f\"ur Nanotechnologie, Karlsruhe Institute of Technology, 76021 Karlsruhe, Germany
}
\affiliation{
 Institut f\"ur Theorie der kondensierten Materie, Karlsruhe Institute of Technology, 76128 Karlsruhe, Germany
}
\affiliation{
 Petersburg Nuclear Physics Institute, 188350 St.\ Petersburg, Russia.
}

\author{M.\ A.\ Skvortsov}
\affiliation{
 L.\ D.\ Landau Institute for Theoretical Physics RAS, 119334 Moscow, Russia
}
\affiliation{
 Moscow Institute of Physics and Technology, 141700 Moscow, Russia
}
\affiliation{
 Skolkovo Institute of Science and Technology, 143025 Skolkovo, Russia
}

\begin{abstract}
We study quantum interference effects in a two-dimensional chiral metal (bipartite lattice) with vacancies. We demonstrate that randomly distributed vacancies 
constitute a peculiar type of chiral disorder leading to strong modifications of critical properties at zero energy as compared to conventional chiral metals. 
In particular, the average density of states diverges as $\rho \propto E^{-1} |\ln E|^{-3/2}$ and the correlation length $L_c \propto \sqrt{|\ln E|}$ in the 
limit $E \to 0$. When the average density of vacancies is different in the two sublattices, a finite concentration of zero modes emerges and a gap in the 
quasiclassical density of states opens around zero energy. Interference effects smear this gap resulting in exponentially small tails at low energies.
\end{abstract}

\maketitle

\textit{1. Introduction.---}
Anderson localization \cite{Anderson58} remains in the focus of condensed matter research. Development of the symmetry \cite{Zirnbauer96} and topology 
\cite{SchnyderKitaev} classification of disordered systems is one of central advances in the field. It has been realized that underlying symmetries and 
topologies induce a rich variety of localization phenomena, including, in particular, critical phases and quantum phase transitions between metallic and 
insulating states \cite{Evers08}.

The symmetry classification of disordered systems \cite{Zirnbauer96, Evers08} includes three families of symmetry classes: conventional (Wigner-Dyson), chiral, 
and superconducting (Bogoliubov-de Gennes). In this paper we consider two-dimensional (2D) chiral models. Chiral symmetry (classes AIII, BDI, and CII) implies 
that the Hamiltonian can be arranged in the form of a block off-diagonal matrix. A standard realization of such a system is provided by a bipartite lattice with
random hopping. In contrast to Wigner-Dyson classes, chiral systems exhibit very unusual localization properties. A remarkable feature of the chiral metal is 
the exact absence of localization corrections to all orders in the perturbation theory \cite{GadeWegner}. At the same time, the density of states (DOS) is 
strongly modified by the quantum interference effects and diverges at the center of the band. As was shown in Ref.\ \onlinecite{Koenig12}, localization effects
do emerge in chiral models when the theory is treated non-perturbatively. Specifically, the localization is controlled by topological vortex-like excitations
of the sigma model, in similarity with the Berezinskii-Kosterlitz-Thouless phase transition \cite{BKT}.

The experimental discovery of graphene \cite{graphene} and extensive study of its peculiar transport properties near the Dirac point has given an additional 
boost to studying quantum transport in systems with chiral symmetry. In particular, long-range lattice corrugations (ripples) in graphene generate an effective 
random magnetic field acting within each valley \cite{MorpurgoGuinea06}, placing the system into the chiral unitary class AIII. A hexagonal lattice with 
vacancies falls into the chiral orthogonal class BDI. A vacancy can be modeled by cutting all lattice bonds adjacent to the vacant site, which yields a special 
type of bond disorder \cite{graphene_vacancies}. Chemical adsorbents, such as hydrogen, attached to individual graphene atoms can be approximated as vacancies 
since the strong on-site potential prevents an electron from occupying the impurity site.

Bipartite lattices with randomly located vacancies belonging to the chiral symmetry class constitute the subject of the present paper. We will show that 
vacancies crucially modify interference effects close to the center of the band (Dirac point in case of graphene) leading to enhanced DOS and reduced 
localization length as compared to other realizations of chiral systems. These modifications are intimately related to zero modes arising in bipartite systems 
with unequal number of sites in the two sublattices. We will develop the non-linear sigma model formalism for chiral systems with vacancies and demonstrate how 
the zero modes affect localization phenomena.

\textit{2. Model and field theory.---}
As a model system, we consider a nearest-neighbor hopping Hamiltonian on the bilayer square lattice with $t$ and $t' + h(\mathbf{r})$ being the intra- and 
inter-layer hopping amplitudes, respectively. The latter amplitude contains a relatively small complex random part $h(\mathbf{r})$. This model belongs to the 
chiral unitary class AIII and is equivalent to the model studied by Gade and Wegner \cite{GadeWegner}. The vacancies will be introduced as a strong 
potential $V$ (to be later sent to infinity) applied to randomly selected sites with the average densities $n_A$ and $n_B$ in the two 
sublattices. The Hamiltonian for such a model has the following form in the sublattice basis:
\begin{equation}
 H
  = \begin{pmatrix}
      V_A(\mathbf{r}) & \xi(\mathbf{p}) + h(\mathbf{r}) \\
      \xi(\mathbf{p}) + h^*(\mathbf{r}) & V_B(\mathbf{r})
    \end{pmatrix}.
 \label{ham}
\end{equation}
We assume that $t'$ is slightly less than $2t$; then the low-energy states are located close to the center of the Brillouin zone and the kinetic part of the 
Hamiltonian acquires the quadratic form $\xi(\mathbf{p}) = p^2/2m - \mu$ with $m = 1/ta^2$ and $\mu = 2t - t'$, where $a$ is the lattice spacing. We assume that 
the random-hopping part obeys a Gaussian distribution with $\langle h \rangle = 0$ and $\langle h(\mathbf{r}) h^*(\mathbf{r}') \rangle = \delta_{\mathbf{r,r'}} 
/ 2\pi\nu\tau$, where $\nu = m/2\pi$ is the clean-system DOS per band and $\tau$ is the mean free time induced by the bond disorder. The on-site potential 
modeling vacancies is contained in the $V_{A,B}$ terms of the Hamiltonian (\ref{ham}). These terms are diagonal in sublattice space and manifestly violate the 
chiral symmetry. However, the symmetry will be restored later when we take the limit $V \to \infty$.

To compute the DOS at low energies, we will use the non-linear sigma model formalism. The starting point of the sigma-model derivation is the replicated action 
for the fermion fields:
\begin{equation}
 S
  = -i \int d\mathbf{r}\, \psi^\dagger \bigl[ E + i0 - H \bigr] \psi.
 \label{Spsi}
\end{equation}
Here $\psi = \{\psi_A, \psi_B\}^T$ is a $2N$-component vector of Grassmann fields operating in the space of two sublattices (A and B) and $N$ replicas. For the
moment, we assume that the Gaussian bond disorder dominates, i.e., the mean free path due to random hopping terms is much shorter than the characteristic
distance between vacancies. Averaging with respect to $h(\mathbf{r})$ gives rise to the spatially local quartic term in the action with the density
$-\mathop{\mathrm{tr}} (\psi_A \psi^\dagger_A \psi_B \psi^\dagger_B)/2\pi\nu\tau$. This term is decoupled by the Hubbard-Stratonovich transformation using an 
auxiliary complex-valued matrix field $Q$ of size $N$. Next, the fermion fields $\psi$ are integrated out, yielding \cite{Altland02}
\begin{equation}
 S
  = \mathop\mathbf{Tr}\! \left[ \frac{\pi\nu}{2\tau}\, Q^\dagger Q
      -\ln \begin{pmatrix}
        E - V_A + \dfrac{iQ}{2\tau}\! &\! -\xi \\
        -\xi \!&\! E - V_B + \dfrac{iQ^\dagger}{2\tau}
      \end{pmatrix}
    \!\right]\!,
 \label{trlog}
\end{equation}
where `\textbf{Tr}' denotes the full operator trace including replica and real space.

In the limit $E = V_{A,B} = 0$, the action (\ref{trlog}) is minimized at the manifold $Q^\dagger Q = 1$. This defines the target space of the sigma model, $Q 
\in \mathrm{U}(N)$ for the symmetry class AIII. With the matrix $Q$ restricted to this target manifold, we expand Eq.\ (\ref{trlog}) up to second order in 
gradients, linear order in $E$, and to \emph{all} orders in $V_{A,B}$. This yields the action of the sigma model $S = S_\sigma + S_E + S_V$ with
\begin{subequations}
\begin{align}
 S_\sigma
  &= \int \frac{d\mathbf{r}}{8\pi} \left[
      \sigma \mathop{\mathrm{tr}} \bigl( \nabla Q^\dagger \nabla Q \bigr)
      - c\, \bigl( \mathop{\mathrm{tr}} Q^\dagger \nabla Q \bigr)^2
    \right], \label{Ssigma} \\
 S_E
  &= i \pi \nu E \int d\mathbf{r} \mathop{\mathrm{tr}} \bigl( Q^\dagger + Q \bigr), \\
 S_V
  &= -\int \frac{d\mathbf{r}}{a^2} \mathop{\mathrm{tr}} \ln \bigl[ 1 + i \pi \nu a^2 \bigl( V_A Q^\dagger + V_B Q \bigr) \bigr].
\end{align}
\end{subequations}
In the kinetic part of the action $S_\sigma$, the parameter $\sigma = 2\pi^2 \nu v^2 \tau$ is the dimensionless Drude conductivity in units $e^2/\pi h$, where
$v$ is the Fermi velocity. The second term in $S_\sigma$, commonly referred to as the Gade term, does not result from the gradient expansion of Eq.\ 
(\ref{trlog}) (i.e., the bare value of $c$ is negligible) but is generated in course of renormalization. In deriving the last part of the action $S_V$, we
assume that the on-site impurities (vacancies) are located far from each other as compared to the mean free path, and hence neglected correlations in the $Q$ 
field at different impurity positions.

We average the action over the Poisson distribution of vacancies and take the limit $V \to \infty$. This results in
\begin{equation}
 S_V
  = \int d\mathbf{r}\, \left[
      n_A \bigl( 1 - \det Q^\dagger \bigr) + n_B \bigl( 1 - \det Q \bigr)
    \right].
\end{equation}
Here $n_{A,B}$ are concentrations of vacancies in the two sublattices. We will further simplify $S_V$ by expanding it in powers of $\ln \det Q$ up to the
second order. Since $\ln \det Q$ scales linearly with the number of replicas $N$, higher terms of such an expansion have no effect in the replica limit 
\cite{footnote_replica_limit}. The final expression reads
\begin{gather}
 S_V
  = \int d\mathbf{r}\, \left[
      2\pi\nu\Delta \mathop{\mathrm{tr}} \ln Q - \frac{n}{2} \bigl( \mathop{\mathrm{tr}} \ln Q \bigr)^2
    \right], \label{SV} \\
 \Delta
  =  (n_A - n_B)/2\pi\nu,
 \qquad
 n
  = n_A + n_B.
\end{gather}
Here we introduced a parameter $\Delta$ that plays the role of the quasiclassical gap in the spectrum, see below.

The assumption $n_{A,B} l^2 \ll 1$ used in our sigma-model derivation is actually not essential. One can consider the extreme case when vacancies are the only 
type of disorder in the system. The mean free path $l = \pi \nu v/n$ is then much longer than the distance between impurities. The sigma model can still be 
derived in this limit with a help of superbosonization technique \cite{superbosonization}; this will be a subject of a separate publication. The only 
modification to the above result is a different numerical factor in the last term of Eq.\ (\ref{SV}), which is of no importance for the analysis below. Finally, 
the average DOS is given, within the replica sigma-model formalism, by
\begin{equation}
 \rho(E)
  = -\mathop{\mathrm{Im}} \lim_{N \to 0} \frac{1}{\pi N} \frac{\partial}{\partial E} \int DQ\, e^{-S[Q]}.
 \label{DOS}
\end{equation}

A comment on the action symmetry is in order here. The kinetic action $S_\sigma$ is invariant under global left and right rotations $Q \mapsto U_L^\dagger Q 
U_R$ with any spatially constant unitary matrices $U_{L,R}$. This symmetry is inherited from the original fermionic action (\ref{Spsi}). Indeed, the latter is 
invariant under the transformation $\psi_{A,B} \mapsto U_{L,R} \psi_{A,B}$, $\psi_{A,B}^\dagger \mapsto \psi_{A,B}^\dagger U_{R,L}^\dagger$, when both energy 
$E$ and on-site potentials $V_{A,B}$ vanish. Although the vacancies preserve the chiral symmetry of the Hamiltonian, the action $S_V$ partially breaks the 
$\mathrm{U}(N) \times \mathrm{U}(N)$ symmetry of the sigma model for the following reason. When the number of vacant sites in the two sublattices is  
different, the number of $\psi_A$ and $\psi_B$ fields is also different. Even though the action (\ref{Spsi}) retains its full symmetry, the invariance of the
corresponding path integral in $\psi$ and $\psi^\dagger$ requires also that the transformation Jacobian is unity, i.e., $\det U_L = \det U_R$. Hence only the
transformations preserving $\det Q$ are true symmetries of the path integral and of the action $S_V$. We conclude that vacancies effectively reduce the target
space of the sigma model from $\mathrm{U}(N)$ down to $\mathrm{SU}(N)$.

\textit{3. Zero-dimensional limit.---}
We begin our analysis of the sigma model with the zero-dimensional (0D) limit. Assume a sample of a finite size $L^2$ and energies low enough to neglect the
kinetic action $S_\sigma$. On average, such sample contains $N_{A,B} = n_{A,B} L^2$ vacancies in the two sublattices. The problem of the spectrum of a random
chiral matrix with the imbalance between A and B states was considered in Refs.\ \onlinecite{Verbaarschot99, Ivanov02}. It was shown that a fixed imbalance 
leads to an additional term in the sigma-model action of the form $(N_A - N_B) \ln \det Q$. This is exactly the first term of Eq.\ (\ref{SV}). The second term 
of $S_V$ describes small Gaussian fluctuations of the imbalance.

Refs.\ \onlinecite{Verbaarschot99, Ivanov02} provide an exact solution of the 0D problem determined by an integral over the whole sigma-model manifold. For our 
purposes an approximate quasiclassical solution valid in the limit $N_{A,B} \gg 1$ is sufficient. To obtain it, we take the spatially constant and
diagonal-in-replicas minimum of the action. Using the ansatz $Q = e^{i \phi}$, we compute the variation of $S_E + S_V$ and obtain, in the limit $N \to 0$,  
$\sin \phi = \Delta/E$. The DOS, Eq.\ (\ref{DOS}), is then given by 
\begin{equation}
 \rho(E)
  = 2\nu \left[
      \pi\, \delta ( E / \Delta) + \sqrt{1 -
\Delta^2/E^2}
    \right].
 \label{DOS_classic}
\end{equation}
The DOS exhibits a gap at energies $|E| < \Delta$ with a delta peak in the center due to zero modes.

\begin{figure}
\includegraphics[width=0.9\columnwidth]{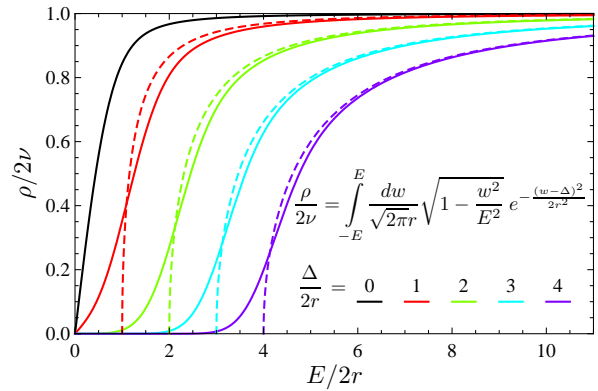}
\caption{(Color online) Average DOS in the 0D chiral system with fluctuating imbalance. Dashed lines show quasiclassical result (\ref{DOS_classic}). At low 
energies DOS vanishes linearly, cf.\ Eq.\ (\ref{0D}).}
\label{fig_0D}
\end{figure}

The fluctuations of the imbalance can be now taken into account by averaging the above result with respect to Gaussian fluctuations of the gap with the mean
value $\Delta$ and dispersion $r = \sqrt{n}/2\pi \nu L$. The result of this averaging is displayed in Fig.\ \ref{fig_0D}. At low energies and at small 
imbalance, $E,\Delta \ll r$, the 0D DOS reads
\begin{equation}
 \rho_\text{0D}(E, L)
  \simeq \sqrt{2\pi}\nu [
      2r \delta(E) + E/r ].
 \label{0D}
\end{equation}

\textit{4. Renormalization group.---}
We are now in a position to solve the 2D problem. The spatial fluctuations of the sigma-model field $Q$ will be taken into account with the help of the
renormalization group (RG). At the last step we will apply the 0D result to the renormalized theory. The RG for the chiral sigma model was first discussed by
Gade and Wegner \cite{GadeWegner}. It was shown that the conductivity is not renormalized to all orders of the perturbation theory in the parameter $1/\sigma 
\ll 1$. At the same time, a new coupling $c$ [see Eq.\ (\ref{Ssigma})] is generated by the RG. The corresponding RG equations have the following form in the 
replica limit: 
\begin{equation}
 \partial\sigma / \partial \ln L
  = 0,
 \qquad
 \partial c / \partial \ln L
  = 1.
 \label{RG_sigma_c}
\end{equation}
Note that both equations are exact to all orders in $1/\sigma$ in the symmetry class AIII. The couplings $\Delta$ and $n$, introduced in the action $S_V$, Eq.\ 
(\ref{SV}), are also not renormalized. To prove this, we separate the matrix $Q = e^{i \phi} U$ into the phase factor and a matrix with unit determinant,
$\det U = 1$. The variables $\phi$ and $U$ decouple at $E = 0$,
\begin{multline}
 S_\sigma + S_V
  = \int d\mathbf{r} \Biggl\{
      \frac{\sigma}{8\pi} \mathop{\mathrm{tr}} \bigl( \nabla U^\dagger \nabla U \bigr) \\
      +N \left[
        \frac{\sigma + N c}{8\pi} \bigl( \nabla \phi \bigr)^2 + 2i \pi \nu \Delta\, \phi + \frac{n N}{2}\, \phi^2
      \right]
    \Biggr\}.
\end{multline}
The action for $\phi$ is quadratic, so its parameters are not renormalized. In particular, this proves that $\sigma$ is not renormalized in the replica limit.
From the above action we also see that the vacancy concentration $n$ indeed provides a mass for the fluctuations of $\phi$ thus breaking the symmetry
$\mathrm{U}(N) \mapsto \mathrm{SU}(N)$ as discussed above.

A finite energy $E$ breaks the chiral symmetry of the problem and couples $\phi$ and $U$ variables. To find the renormalization of energy, we separate fast and
slow fields $Q = Q_\text{slow} Q_\text{fast}$ and expand $Q_\text{fast} = 1 + i W - W^2/2$. The dynamics of $Q_\text{fast}$ is governed by the action $S_\sigma
+ S_V$ yielding the propagator
\begin{equation}
 \langle W_\mathbf{-q} W_\mathbf{q} \rangle
  = \frac{4\pi}{\sigma q^2} \left[
      N - \frac{c q^2 + 4 \pi n}{\sigma q^2 + N(c q^2 + 4 \pi n)}
    \right].
 \label{WW}
\end{equation}
Correction to the energy is represented in the differential form $dE/E = - \langle W^2 \rangle /2$, where the right-hand side is integrated over the fast 
momentum shell $L < q^{-1} < L + dL$. Using Eq.\ (\ref{WW}) and taking the limit $N \to 0$, we obtain the one-loop flow equation for energy:
\begin{equation}
 \partial \ln E / \partial \ln L
  = (c + 4\pi n L^2)/\sigma^2.
 \label{RG_E}
\end{equation}
In the absence of vacancies ($n = 0$), this equation reproduces the result of Ref.\ \onlinecite{GadeWegner}. However, any finite concentration $n$ eventually 
leads to a dramatic acceleration of the energy renormalization. Indeed, the parameter $c$ grows only as $\ln L$ according to Eq.\ (\ref{RG_sigma_c}). Thus we 
will neglect $c$ compared to $n L^2$, which is always justified at long scales (low energies).

The RG flow stops at a critical scale $L_c$ determined by $\sigma/\nu L_c^2 \sim \max\{ \tilde E, \Delta \}$, where $\tilde E \sim E\, e^{n L_c^2/\sigma^2}$ is 
the renormalized energy. The DOS is then given by the 0D result, Eq.\ (\ref{0D}), taken at the scale $L_c$: $\rho(E) = \rho_\text{0D} (\tilde E, L_c) \tilde
E/E$. The factor $\tilde E/E$ appears here due to renormalization of energy in the derivative in Eq.\ (\ref{DOS}). 

In the balanced case $n_A = n_B$, i.e., $\Delta = 0$, we have the following result for the correlation length and DOS:
\begin{gather}
 L_c(E)
  \sim \sigma n^{-1/2} |\ln E \tau_n|^{1/2}, \label{Lc} \\
 \rho(E)
  \sim \frac{\sigma^2}{\sqrt{n} E L_c^3}
  \sim \frac{\nu}{E \tau_n |\ln E \tau_n|^{3/2}}. \label{DOS_balance}
\end{gather}
Here we introduced a time scale related to vacancies, $\tau_n = 4\pi \nu \sigma/n$. 

These results should be contrasted to low-energy behavior found by Gade and Wegner for the chiral model without vacancies, $L_c(E) \sim \exp[\sigma|\ln E
|^\kappa]$ with $\kappa=1/2$ and $\rho(E) \sim 1/ EL_c^2(E)$ \cite{GadeWegner, Motrunich02}. We see that in the presence of vacancies the correlation length 
diverges much more slowly at $E\to 0$ and the DOS develops a stronger singularity. 

\begin{figure}
\includegraphics[width=0.9\columnwidth]{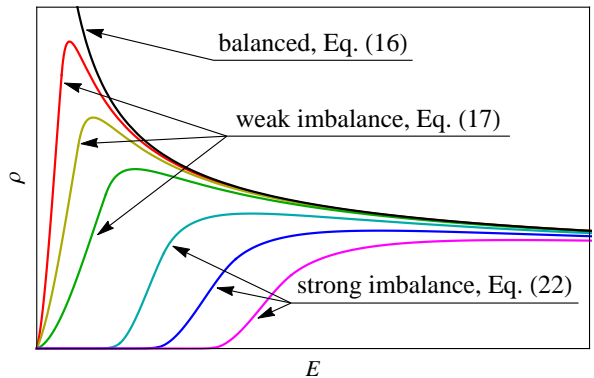}
\caption{(Color online) Schematic average DOS in the 2D chiral metal with vacancies. Black line shows the result (\ref{DOS_balance}) in the limit $\Delta = 0$.
Weak imbalance results in linear decay of DOS at exponentially low energies (\ref{DOS_imbalance}). Strong imbalance leads to the gap with exponentially small
subgap tail (\ref{DOS_tail}).}
\label{fig_2D}
\end{figure}

When vacancies are \textit{weakly imbalanced}, $\Delta \tau_n \ll 1$, the correlation length saturates at the value $L_c = \sqrt{\sigma/\nu \Delta}$ and the 
DOS linearly drops to zero at exponentially small energies. A finite density of zero modes introduces a delta peak in the DOS:
\begin{equation}
 \rho\bigl(E \ll \Delta e^{-1/\Delta\tau_n}\bigr)
  \propto \frac{\nu}{\sqrt{\Delta \tau_n}} \left[
      \delta \left( \frac{E}{\Delta} \right)
      +E\tau_n\, e^{1/\Delta\tau_n}
    \right].
 \label{DOS_imbalance}
\end{equation}
Here the amplitude of the delta peak is accurate up to a numerical factor of order unity while the second term contains a similar factor in the exponent. A 
crossover between Eqs.\ (\ref{DOS_balance}) and (\ref{DOS_imbalance}) in the case of weak imbalance is illustrated in Fig.\ \ref{fig_2D}.

\textit{5. Strong imbalance.---}
When the imbalance is strong, $\Delta\tau_n \gg 1$, renormalization of energy is weak and the DOS is approximately given by the quasiclassical result 
(\ref{DOS_classic}), see Fig.\ \ref{fig_2D}. Rare fluctuations produce an exponentially small subgap tail at $E < \Delta$. We will compute this tail by the 
optimal fluctuation method. In the sigma-model formalism it amounts to finding a saddle point of the action (instanton) breaking the replica symmetry. The 
problem is similar to the calculation of subgap DOS in a superconductor with fluctuating order parameter \cite{LarkinOvchinnikov71, Silva05, Skvortsov13}.

We look for an instanton with diagonal matrix structure $Q = e^{i \phi_{1,2}}$ with the phase $\phi_1$ in one replica and $\phi_2$ in other $N - 1$ replicas.
The variation of the total action $S_\sigma + S_E + S_V$ (with the Gade term neglected) yields two coupled equations for $\phi_{1,2}$:
\begin{gather}
 \frac{\sigma}{4\pi} \nabla^2 \phi_1 - U'(\phi_1)
  = \frac{\sigma}{4\pi} \nabla^2 \phi_2 - U'(\phi_2)
  = n (\phi_1 - \phi_2), \label{instanton}\\
 U(\phi)
  = 2i \pi \nu (\Delta\, \phi + E \cos\phi).
\end{gather}
Assuming that the coupling $n$ is sufficiently strong \cite{footnote_strong_n}, we substitute $\phi_{1,2} = \phi \pm \chi/2$ and expand the equations to linear
order in $\chi$. Next, we exclude $\chi$ and obtain a closed differential equation for $\phi$. We will look for a circular symmetric solution $\phi(r)$.
Remarkably, the fourth-order equation for $\phi$ can be simplified with the help of a dimensional reduction trick \cite{Silva05, Skvortsov13} down to the 
second order,
\begin{equation}
 \frac{\sigma}{4\pi} \left( \frac{\partial^2 \phi}{\partial r^2} - \frac{1}{r} \frac{\partial \phi}{\partial r} \right) - U'(\phi)
  = 0.
 \label{inst_phi}
\end{equation}
The differential part has the form of the ``radial Laplace operator in zero dimensions.'' The transformation leading to Eq.\ (\ref{inst_phi}) is enabled due to
a hidden supersymmetry of the problem \cite{ParisiSourlas79}.

Although the nonlinear equation (\ref{inst_phi}) is not analytically solvable, the fact that the values of $\phi$ at the origin and at infinity satisfy
$U'(\phi_{0,\infty}) = 0$ suffices to compute the instanton action. Expanding the action in $\chi$, we perform the radial integration by using Eq.\ 
(\ref{inst_phi}),
\begin{multline}
 \!\!\!S_\text{inst}
  = -\!\int\! \frac{d\mathbf{r}}{2n} \left[
      \frac{\sigma}{4\pi} \nabla^2 \phi - U'(\phi)
    \right]^2\!
  = \frac{\sigma}{n} \bigl[
      U(\phi_\infty) - U(\phi_0)
    \bigr] \\
  = \Delta \tau_n \left[
      \mathop{\mathrm{arccosh}} (\Delta/E) - \sqrt{1 -
 E^2/\Delta^2}
    \right].
 \label{Sinst}
\end{multline}
This action determines the subgap DOS with exponential accuracy provided $S_\text{inst} \gg 1$ (see Fig.\ \ref{fig_2D}),
\begin{equation}
 \rho
  \propto e^{-S_\text{inst}}
  \propto \begin{cases}
      \exp \left[ -\dfrac{\Delta \tau_n}{3} (2\epsilon)^{3/2} \right], & \epsilon = 1 - \dfrac{E}{\Delta} \ll 1, \\
      \bigl( E/\Delta \bigr)^{\Delta \tau_n}, & E \ll \Delta.
    \end{cases}
 \label{DOS_tail}
\end{equation}
The ``near'' tail at $(\Delta\tau_n)^{-2/3} \ll \epsilon \ll 1$ has the form characteristic for the distribution of large eigenvalues of a random matrix
\cite{TracyWidom}. This effectively 0D result appears in our problem due to dimensional reduction. The deep subgap tail at $E \ll \Delta$ is similar to the 2D
results of Ref.\ \onlinecite{MuzykantskiiKhmelnitskii} for the long-time asymptotics of the current relaxation due to anomalously localized states.

\textit{6. Summary and outlook.---}
To summarize, we have demonstrated that randomly distributed vacancies in a 2D chiral metal strongly modify its critical properties at zero energy. They reduce
the correlation length (\ref{Lc}) and increase the DOS (\ref{DOS_balance}) as compared to the conventional chiral metals \cite{GadeWegner}. Technically, this 
modification occurs due to an additional term in the sigma-model action (\ref{SV}). This term provides a mass to the phase variable $\ln \det Q$ and reduce the 
sigma-model target space $\mathrm{U}(N) \mapsto \mathrm{SU}(N)$ in the symmetry class AIII. Similar reduction will take place in the other chiral classes, BDI 
and CII, leading to the models on the $\mathrm{SU}(2N)/\mathrm{Sp}(2N)$ and $\mathrm{SU}(N)/\mathrm{O}(N)$ manifolds, respectively.

Graphene with vacancies at the Dirac point represents a class BDI system with $\sigma \sim 1$. While our analysis was performed for $\sigma \gg 1$, our results 
(\ref{Lc}), (\ref{DOS_balance}) are in agreement with numerical study of graphene lattice with vacancies \cite{Evers, Evers-theses}. 

When the density of vacancies is different in the two sublattices, $n_A \neq n_B$, the system hosts exact  zero modes with the concentration $|n_A - n_B|$.
A weak imbalance depletes the DOS at low energies linearly, Eq.\ (\ref{DOS_imbalance}), while a strong imbalance opens a gap in the spectrum with an 
exponentially small tail (\ref{DOS_tail}) at low energies. 

The reduction of the sigma-model symmetry by generation of a mass in the $\mathrm{U}(1)$ sector due to vacancies has also profound implications for 
localization properties at chiral-symmetry point ($E=0$). Indeed, topological vortex excitations of the U(1) phase $\ln \det Q$ are responsible for the
localization in chiral metals at sufficiently strong disorder \cite{Koenig12}. Our results suggest that random vacancies disable this mechanism and thus 
prevent the localization. This is similar to classes D and DIII where the localization can only emerge due to domain walls associated with a discrete 
$\mathrm{O}(1)$ degree of freedom $\det Q = \pm 1$ \cite{gruzberg13}. This degree of freedom gets frozen by vortex impurities \cite{Bocquet00}, which leads to 
disappearance of the localized phase. 

We thank F.\ Evers for sharing numerical data \cite{Evers} prior to publication and for discussions. This work was supported by DFG SPPP 1459 and 1666, 
by GIF, and by BMBF.


\begin{thebibliography}{99}

\bibitem{Anderson58}
P.\ W.\ Anderson, Phys.\ Rev.\ \textbf{109}, 1492 (1958).

\bibitem{Zirnbauer96}
M.\ R.\ Zirnbauer, J.\ Math.\ Phys.\ \textbf{37}, 4986 (1996);
A.\ Altland and M.\ R.\ Zirnbauer, Phys.\ Rev.\ B \textbf{55}, 1142 (1997).

\bibitem{SchnyderKitaev}
A.\ P.\ Schnyder, S.\ Ryu, A.\ Furusaki, and A.\ W.\ W.\ Ludwig, Phys.\ Rev.\ B \textbf{78}, 195125 (2008); AIP Conf.\ Proc.\ \textbf{1134}, 10 (2009);
A.\ Kitaev, \textit{ibid.} \textbf{1134}, 22 (2009).

\bibitem{Evers08}
F.\ Evers and A.\ D.\ Mirlin, Rev.\ Mod.\ Phys.\ \textbf{80}, 1355 (2008). 

\bibitem{GadeWegner}
R.\ Gade and F.\ Wegner, Nucl.\ Phys.\ B \textbf{360}, 213 (1991).

\bibitem{Koenig12}
E.\ J.\ K\"onig, P.\ M.\ Ostrovsky, I.\ V.\ Protopopov, and A.\ D.\ Mirlin, Phys.\ Rev.\ B \textbf{85}, 195130 (2012).

\bibitem{BKT}
V.\ L.\ Berezinskii, Zh.\ Eksp.\ Teor.\ Fiz.\ \textbf{61}, 1144 (1971) [Sov.\ Phys.\ JETP \textbf{34}, 610 (1972)];
J.\ M.\ Kosterlitz and D.\ J.\ Thouless, J.\ Phys.\ C \textbf{5}, L124 (1972); \textbf{6}, 1181 (1973).

\bibitem{graphene}
A.\ H.\ Castro Neto, F.\ Guinea, N.\ M.\ R.\ Peres, K.\ S.\ Novoselov, and A.\ K.\ Geim, Rev.\ Mod.\ Phys.\ \textbf{81}, 109 (2009). 

\bibitem{MorpurgoGuinea06}
A.\ F.\ Morpurgo and F.\ Guinea, Phys.\ Rev.\ Lett.\ \textbf{97}, 196804 (2006).

\bibitem{graphene_vacancies}
M.\ Titov, P.\ M.\ Ostrovsky, I.\ V.\ Gornyi, A.\ Schuessler, and A.\ D.\ Mirlin, Phys.\ Rev.\ Lett.\ \textbf{104}, 076802 (2010);
P.\ M.\ Ostrovsky, M.\ Titov, S.\ Bera, I.\ V.\ Gornyi, and A.\ D.\ Mirlin, \textit{ibid.} \textbf{105}, 266803 (2010);
S.\ Gattenl\"ohner, W.-R.\ Hannes, P.\ M.\ Ostrovsky, I.\ V.\ Gornyi, A.\ D.\ Mirlin, and M.\ Titov, \textit{ibid.} \textbf{112}, 026802 (2014).

\bibitem{Altland02}
A.\ Altland, Phys.\ Rev.\ B \textbf{65}, 104525 (2002).

\bibitem{footnote_replica_limit}
More precisely, neglecting higher terms in the expansion in $\mathop{\mathrm{tr}} \ln Q$ is justified due to weak replica symmetry breaking. The RG,
Eqs.\ (\ref{RG_sigma_c}), (\ref{RG_E}), accounts for fluctuations around replica-symmetric configuration, while the instanton solution in the imbalanced case
strongly breaks replica symmetry only at extremely low energies \cite{footnote_strong_n}.

\bibitem{superbosonization}
K.\ B.\ Efetov, G.\ Schwiete, and K.\ Takahashi, Phys.\ Rev.\ Lett.\ \textbf{92}, 026807 (2004);
J.\ E.\ Bunder, K.\ B.\ Efetov, V.\ E.\ Kravtsov, O.\ M.\ Yevtushenko, and M.\ R.\ Zirnbauer, J.\ Stat.\ Phys.\ \textbf{129}, 809 (2007);
P.\ Littelmann, H.-J.\ Sommers, and M.\ R.\ Zirnbauer, Commun.\ Math.\ Phys.\ \textbf{283}, 343 (2008).

\bibitem{Verbaarschot99}
P.\ H.\ Damgaard, J.\ C.\ Osborn, D.\ Toublan, and J.\ J.\ M.\ Verbaarschot, Nucl.\ Phys.\ B \textbf{547}, 305 (1999).

\bibitem{Ivanov02}
D.\ A.\ Ivanov, J.\ Math.\ Phys.\ \textbf{43}, 126 (2002).

\bibitem{Motrunich02}
As was shown by
O.\ Motrunich, K.\ Damle, and D.\ A.\ Huse, Phys.\ Rev.\ B \textbf{65}, 064206 (2002);
C.\ Mudry, S.\ Ryu, and A.\ Furusaki, Phys.\ Rev.\ B \textbf{67}, 064202 (2003),
fluctuations of DOS modify the exponent $\kappa$: $1/2 \to 2/3$. 

\bibitem{LarkinOvchinnikov71}
A.\ I.\ Larkin and Yu.\ N.\ Ovchinnikov, Zh.\ Eksp.\ Teor.\ Fiz.\ \textbf{61}, 2147 (1971) [Sov.\ Phys.\ JETP \textbf{34}, 1144 (1972)].

\bibitem{Silva05}
A.\ Silva and L.\ B.\ Ioffe, Phys.\ Rev.\ B \textbf{71}, 104502 (2005).

\bibitem{Skvortsov13}
M.\ A.\ Skvortsov and M.\ V.\ Feigel'man, Zh.\ Eksp.\ Teor.\ Fiz.\ \textbf{144}, 560 (2013) [JETP \textbf{117}, 487 (2013)].

\bibitem{footnote_strong_n}
Actual validity conditions for the expansion in $\phi_1 - \phi_2$ depend on the energy. Close to the edge of the gap, $\epsilon = 1 - E/\Delta \ll 1$, the
expansion is controlled by the inequality $n \gtrsim \nu\Delta \gg \nu\Delta\sqrt{\epsilon}$ and hence is always valid. At small energies, $E \ll \Delta$, the
condition is extremely weak: $n \gg \nu \Delta \ln \ln (\Delta/E)$. It breaks down at overwhelmingly low energies provided the imbalance is small, $n \gg
\nu\Delta$.

\bibitem{ParisiSourlas79}
G.\ Parisi and N.\ Sourlas, Phys.\ Rev.\ Lett.\ \textbf{43}, 744 (1979).

\bibitem{TracyWidom}
C.\ A.\ Tracy and H.\ Widom, Commun.\ Math.\ Phys.\ \textbf{159}, 151 (1994); \textbf{177}, 727 (1996).

\bibitem{MuzykantskiiKhmelnitskii}
B.\ A.\ Muzykantskii and D.\ E.\ Khmelnitskii, Phys.\ Rev.\ B \textbf{51}, 5480 (1995).

\bibitem{Evers}
V.\ H\"afner, J.\ Schindler, N.\ Weik, T.\ Mayer, S.\ Balakrishnan, R.\ Narayanan, S.\ Bera,  F.\ Evers, arXiv:1404.6138.

\bibitem{Evers-theses}
V.\ H\"afner, Diploma thesis, Karlsruhe Institute of Technology, 2011;
J.\ Schindler, Diploma thesis, Karlsruhe Institute of Technology, 2012.

\bibitem{gruzberg13}
I.\ A.\ Gruzberg, A.\ D.\ Mirlin, and M.\ R.\ Zirnbauer, Phys.\ Rev.\ B \textbf{87}, 125144 (2013). 

\bibitem{Bocquet00}
N.\ Read and D.\ Green, Phys.\ Rev.\ B \textbf{61}, 10267 (2000); 
M.\ Bocquet, D.\ Serban, and M.\ R.\ Zirnbauer, Nucl.\ Phys.\ B \textbf{578}, 628 (2000);
J.\ T.\ Chalker, N.\ Read, V.\ Kagalovsky, B.\ Horovitz, Y.\ Avishai, and A.\ W.\ W.\ Ludwig, Phys.\ Rev.\ B \textbf{65}, 012506 (2001). 

\end{thebibliography}
\end{document}